\documentclass[12pt,preprint]{aastex}
\usepackage{epsf,color}

\newcommand{\sfrac}[2]{\mathchoice
  {\kern0em\raise.5ex\hbox{\the\scriptfont0 #1}\kern-.15em/
   \kern-.15em\lower.25ex\hbox{\the\scriptfont0 #2}}
  {\kern0em\raise.5ex\hbox{\the\scriptfont0 #1}\kern-.15em/
   \kern-.15em\lower.25ex\hbox{\the\scriptfont0 #2}}
  {\kern0em\raise.5ex\hbox{\the\scriptscriptfont0 #1}\kern-.2em/
   \kern-.15em\lower.25ex\hbox{\the\scriptscriptfont0 #2}}
  {#1\!/#2}}

\newcommand{\myhalf}{\sfrac{1}{2}}
\newcommand{\nph}{{n + \myhalf}}
\newcommand{\nmh}{{n - \myhalf}}

\newcommand{\omegadot}{\dot{\omega}}
\newcommand{\er}{\mathbf{e}_r}

\newcommand{\x}{{\bf x}}

\newcommand{\ubold}{U}
\newcommand{\Abar}{\bar{A}}

\begin{document}

\title{Low Mach Number Modeling of Type Ia Supernovae}

\shorttitle{Low Mach Number Modeling of SNe Ia}
\shortauthors{Almgren et al.}

\author{A.~S.~Almgren\altaffilmark{1},
        J.~B.~Bell\altaffilmark{1},
        C.~A.~Rendleman\altaffilmark{1},
        M.~Zingale\altaffilmark{2}}

\altaffiltext{1}{Center for Computational Science and Engineering,
                 Lawrence Berkeley National Laboratory,
                 Berkeley, CA 94720}

\altaffiltext{2}{Dept. of Astronomy \& Astrophysics,
                 The University of California, Santa Cruz,
                 Santa Cruz, CA 95064}

\begin{abstract}

We introduce a low Mach number equation set for the large-scale
numerical simulation of carbon-oxygen white dwarfs experiencing
a thermonuclear deflagration. Since most of the interesting physics in a
Type Ia supernova transpires at Mach numbers from 0.01 to 0.1, such an
approach enables both a considerable increase in accuracy and savings
in computer time compared with frequently used compressible codes.
Our equation set is derived from the fully compressible equations
using 
low Mach number asymptotics, but without any restriction on the
size of perturbations in density or temperature.  Comparisons with
simulations that use the 
fully compressible equations validate the low
Mach number model in regimes where both are applicable.  Comparisons
to simulations based on the more traditional anelastic approximation
also demonstrate the agreement of these models in the regime for which
the anelastic approximation is valid. 
For low Mach number flows with potentially finite amplitude variations
in density
and temperature, the low Mach number model overcomes the limitations
of each of the more traditional models and can serve as the basis for
an accurate and efficient simulation tool.

\end{abstract}


\section{Introduction}

A broad range of interesting phenomena in science and engineering
occur in a low Mach number regime in which the fluid velocity is much
less than the speed of sound. Several low Mach number schemes have
been developed to exploit this separation of scales;  these models
capture the fluid dynamics of interest without the need to resolve
acoustic wave propagation. Physically, one can think of the solution
to a low Mach number model as supporting infinitely fast acoustic
equilibration rather than finite-velocity acoustic wave propagation.
Mathematically, this is manifest in the addition of a constraint on the 
velocity field to the system of evolution equations. This velocity
constraint can be translated into an elliptic equation for pressure
that expresses the equilibration process.  Because 
explicit discretization schemes for the low Mach number system are limited by
the fluid velocity and not by the sound speed, they often gain
several orders of magnitude in computational efficiency over 
the traditional compressible approach.

The simplest low Mach number model is expressed by the
incompressible Navier-Stokes equations for a constant density fluid.
Generalizations 
that incorporate variations in density include the Boussinesq 
approximation~\citep{Bou03}, which allows heating-induced buoyancy in a 
constant density background, and the anelastic
atmospheric~\citep{Bat53,OguPhi62,DutFic69,Gou69,LipHem82,LipHem85,Lip90,WilOgu72}
and stellar~\citep{LatSpiTooZah76,GilGla81:1,Glatzmaier:1984} approximations that include the 
effect of large-scale background stratification in the fluid density 
and pressure but assume small thermodynamic perturbations from the 
background.  Low Mach number models for chemical
combustion~\citep{rehmBaum:1978,majdaSethian:1985,DayBell:2000} and 
nuclear burning~\citep{Bell:2004} incorporate large compressibility 
effects due to chemical\slash nuclear reactions and thermal processes with a
spatially constant background pressure.

Low Mach number models to date have, with one exception, allowed
either zero volumetric changes (incompressible Navier-Stokes,
Boussinesq) or changes due only to local heating effects (low-speed
combustion, nuclear burning), or to large-scale background
stratification (anelastic).  The only low Mach number model that
incorporates both finite local expansion due to heating and volume
changes due to background stratification is the pseudo-incompressible
equation set for the terrestrial atmosphere,  
introduced by \citet{durran:1989} and rigorously derived using
low Mach number asymptotics by~\citet{bottaKleinAlmgren:1999}.
The formulation of the pseudo-incompressible constraint assumes
the ideal gas equation of state, which results in a simplification 
of terms that does not hold for more general 
equations of state and non-trivial changes in composition.

Our anticipated application for this model is the convective, ignition, 
and early propagation phases of Type~Ia supernovae, but the model should be applicable 
to many other problems, such as Type~I X-ray bursts (see, e.g., 
recent work by \citet{Lin:2005} for an alternate form of the low Mach 
number approach for Type~I X-ray bursts), classical novae, and 
ordinary convection in stars.  Events such as Type~Ia supernovae are 
characterized by a large range in length scales, from the $O(10^{-4})$ 
to $O(10^1)$~cm scale of the flame to the $O(10^8)$~cm scale of the white 
dwarf.  The range in timescales is equally impressive, from the 
100~years of convection that precedes ignition to the one second duration 
of the explosion.  Presently, most large-scale calculations focus on 
the explosion itself, beginning with several seeded hot spots to begin 
the runaway.  In the last minutes of the convective phase, velocities 
reach $\sim 1\%$ of the sound speed 
\citep{woosley2001,woosleywunschkuhlen2004}, with temperature 
fluctuations of at most 5\% \citep{wunschwoosley2004}.  These speeds 
are too slow for compressible codes to accurately follow.  For this 
reason, simulation of the convective and ignition phases has seen only 
limited numerical work, e.g.~\citet{hoflichstein:2002}.  Recent 
analytic work \citep{woosleywunschkuhlen2004} suggests that full star 
simulations are needed to accurately capture the convective flows and 
yield the spatial, temporal, and size distribution of the hot spots 
that seed the explosion.  
Three-dimensional anelastic calculations \citep{kuhlen-ignition:2005} have 
shown that a dipole velocity field dominates the convection.

The low Mach number model presented here, like the anelastic model, 
will be capable of the long time integration necessary to follow the convection.
Unlike the anelastic model, however, the low Mach number approach
continues to be valid as the variation in density and temperature
increases in the flame bubbles that evolve in the early phases
of the explosion.  Following the evolution from the convection 
through the early phases of the explosion is the eventual target of 
our low Mach number methodology.

Although the derivation of the low Mach number equation set 
will be general, the example we will 
consider is a simplified problem without reactions or thermal conduction, 
and with a time-independent radially symmetric form of self-gravity.  
The focus of the examples is to demonstrate 
the ability of the low Mach number model to accurately represent 
the hydrodynamics.  We will compare the simulations based on the low Mach 
number approach to simulations based on the fully compressible equation set, 
where applicable, and to the traditional anelastic approach, where it is 
applicable.  We will show that the low Mach number algorithm works well for very
low Mach number flows, with validation presented up to Mach numbers 
$\sim 0.2$.

In the next section we derive the low Mach number equations, and in
Section~\ref{sec:numer} we discuss the numerical implementation.
Section~\ref{sec:compare} contains
numerical comparisons with compressible and anelastic simulations, and
in the final section we discuss our conclusions and future work.

\section{Low Mach Number Model}

We begin with the fully compressible equations governing motion
in the stellar environment as described, for example, in~\citet{Bell:2004}
\begin{eqnarray}
        \frac{\partial \rho}{\partial t} + \nabla \cdot (\rho \ubold) &=& 0 \enskip , \label{rho_eq_full} \\
        \frac{\partial \rho \ubold}{\partial t} + \nabla \cdot (\rho \ubold \ubold)
         + \nabla p &=& -\rho g \er \enskip  \label{mom_eq_full} , \\
        \frac{\partial \rho E}{\partial t} + \nabla \cdot (\rho \ubold E + p \ubold) &=&  
        \nabla \cdot (\kappa \nabla T) - \rho g (\ubold \cdot \er) - 
        \sum_k {\rho q_k \omegadot_k} \enskip , \label{energy_eq_full} \\
        \frac{\partial \rho X_k}{\partial t} + \nabla \cdot (\rho \ubold X_k) &=&  
        \rho \omegadot_k  \enskip \label{species_eq_full} .
\end{eqnarray}
Here $\rho$, $\ubold$, $T$, and $p$ are the density, velocity,
temperature and pressure, respectively, and $E = e + \ubold \cdot
\ubold / 2$ is the total energy with $e$ representing the internal
energy.  In addition, $X_k$ is the abundance of the $k$th isotope,
with associated production rate, $\omegadot_k$, and
energy release, $q_k$. Finally, $g(r)$ is the radially dependent
gravitational acceleration (resulting from spherically symmetric self-gravity), 
$\er$ is the unit vector in the radial direction, and $\kappa$ is the thermal
conductivity.  The Reynolds number of flows in a typical white dwarf 
is sufficiently large that we neglect viscosity here, though viscous terms
could easily be included in the model and the numerical methodology.

For the stellar conditions being considered the pressure contains
contributions from ions, radiation, and electrons.  Thus
\begin{equation}
 p = p(\rho,T,X_k) = p_\mathrm{ion} + p_\mathrm{rad}  + p_\mathrm{ele} \enskip ,
\label{eq:eos}
\end{equation}
where
\[
  p_\mathrm{ion} = \frac{\rho k_B T}{\Abar m_p} \enskip , \enskip\enskip  p_\mathrm{rad} = a T^4 / 3 \enskip ,
\]
and $p_\mathrm{ele}$ is the contribution to the thermodynamic pressure due to
fermions.  In these expressions $m_p$ is the mass of the proton, $a$
is related to the Stefan-Boltzmann constant $\sigma = a c / 4$, $c$ is
the speed of light, $\Abar = \sum_k {X_k A_k}$, $A_k$ is the atomic
number of the $k$th isotope, and $k_B$ is Boltzmann's constant.  
The ionic
component has the form associated with an ideal gas but the radiation
and electron pressure components do not.  We use a stellar equation of state
as implemented in~\citet{timmes_swesty:2000}.

As a prelude to developing the low Mach number equations, we first
rewrite the energy equation (Eq.~[\ref{energy_eq_full}]) in terms of the enthalpy, $h = e + p/\rho$,
\begin{eqnarray}
\rho \frac{Dh}{Dt} - \frac{Dp}{Dt}
   &=& \nabla \cdot (\kappa \nabla T) - \sum_k\rho q_k \omegadot_k = \rho H \label{enthalpy_eq} \enskip ,
\end{eqnarray}
where we introduce $H$ to represent the enthalpy source terms.


Our goal in this section is to derive a model
for low speed flows in a hydrostatically balanced, 
radially stratified, background 
that removes acoustic waves yet allows for the development of finite amplitude temperature 
and density variations.  We thus posit the existence of a background
state with pressure, density and temperature $p_0(r,t), \rho_0(r,t)$ and
$T_0(r,t)$ satisfying both the equation of state and hydrostatic equilibrium.
Because we will neglect reaction terms
that could potentially alter the large-scale pressure distribution
within the star, for the purposes of this paper we will 
neglect time variation of the background state, i.e., we will assume 
$\partial p_0/\partial t = \partial \rho_0/\partial t = \partial T_0/\partial t = 0$.

In order to understand the behavior of the system, we examine the balance
of terms as a function of Mach number, $M = U/c_s$ 
($c_s$ is the speed of sound), which will be assumed to be small. 
We re-write the momentum equation in 
nondimensional coordinates, where the space and time coordinates as well as the density, 
velocity and pressure, are scaled by characteristic values 
$L_\mathrm{ref}$, $t_\mathrm{ref},$ $\rho_\mathrm{ref}, p_\mathrm{ref},$ and $U_\mathrm{ref}$, respectively.  
For the problem scale of interest $U_\mathrm{ref}$ is a typical advective velocity, 
$L_\mathrm{ref}$ a typical length scale,
$t_\mathrm{ref} = L_\mathrm{ref} / U_\mathrm{ref}$, and $p_\mathrm{ref} = \rho_\mathrm{ref} \; c_{s_\mathrm{ref}}^2$, where
$c_{s_\mathrm{ref}}$ is a characteristic value of $c_s$.
We define a scaling for $g$ in terms of the pressure scale height,
$H_\mathrm{ref} = p_\mathrm{ref} / (\rho_\mathrm{ref} \; g).$

The momentum equation  in nondimensional coordinates ($\tilde{t} = t / t_\mathrm{ref}$, etc.),
and exploiting hydrostatic equilibrium of the reference state, has the form 
\[
 \frac{\partial \tilde{\rho} \tilde{\ubold}}{\partial \tilde{t}} + 
\tilde{\nabla} \cdot (\tilde{\rho} \tilde{\ubold} \tilde{\ubold}) 
+ \frac{1}{M^2} \tilde{\nabla} (\tilde{p} - \tilde{p}_0)
= - \frac{1}{M^2} \frac{L_\mathrm{ref}}{H_\mathrm{ref}} (\tilde{\rho}-\tilde{\rho_0}) \tilde{g} \er \enskip ,
\]
For the large-scale near-equilibrium behavior, we set $L_\mathrm{ref} = H_\mathrm{ref}$,
getting
\[
 \frac{\partial \tilde{\rho} \tilde{\ubold}}{\partial \tilde{t}} + 
\tilde{\nabla} \cdot (\tilde{\rho} \tilde{\ubold} \tilde{\ubold}) 
+ \frac{1}{M^2} \tilde{\nabla} (\tilde{p} - \tilde{p}_0)
= - \frac{1}{M^2} (\tilde{\rho}-\tilde{\rho_0}) \tilde{g} \er \enskip ,
\]
where $\tilde{g} = g \; / \; (p_\mathrm{ref} / (\rho_\mathrm{ref} H_\mathrm{ref})).$ 

Since all nondimensional terms are $O(1),$ it is clear that to maintain
a long-term balance, both $(\tilde{p} - \tilde{p}_0)$ and 
$(\tilde{\rho}-\tilde{\rho_0})$ must be $O(M^2).$   This is consistent
with the traditional anelastic approximation; once the density perturbation
is assumed small the approximation $\nabla \cdot (\rho_0 \ubold) = 0$
follows from the continuity equation, and a linearized temperature-density
relationship can be used to replace the buoyancy term in the momentum
equation by one dependent on temperature (or entropy) rather than density.

Here, however, we are interested in finite amplitude density perturbations.
In this case, it is possible for the model to breakdown in long time integrations 
should the flow accelerate to the point that $M$ is no longer small.
We would, nevertheless, expect the low Mach number model to remain valid 
for a limited period of time.
The behavior of the model for finite time intervals can be examined by
considering a shorter time scale, 
$t_\mathrm{ref} = U_\mathrm{ref} / g$, defined such that on this time scale the buoyancy forcing 
from finite amplitude density perturbations can accelerate the flow to at most $U_\mathrm{ref}.$  
Then, recalling $t_\mathrm{ref} = L_\mathrm{ref} / U_\mathrm{ref}$, we see that
$L_\mathrm{ref} = U^2_\mathrm{ref} / g.$   Recalling then that 
$H_\mathrm{ref} = p_\mathrm{ref} / (\rho_\mathrm{ref} \; g)$ and
$p_\mathrm{ref} = \rho_\mathrm{ref} \; c_{s_\mathrm{ref}}^2$, 
we see that $L_\mathrm{ref} / H_\mathrm{ref} = O(M^2).$
In this case the nondimensional momentum equation has the form
\[
 \frac{\partial \tilde{\rho} \tilde{\ubold}}{\partial \tilde{t}} + 
\tilde{\nabla} \cdot (\tilde{\rho} \tilde{\ubold} \tilde{\ubold}) 
+ \frac{1}{M^2} \tilde{\nabla} (\tilde{p} - \tilde{p}_0)
= - (\tilde{\rho}-\tilde{\rho_0}) \tilde{g} \er \enskip ,
\]
which is consistent with the low Mach number nuclear burning model used
in~\citet{Bell:2004}.  The assumption that $t_\mathrm{ref} = U_\mathrm{ref} / g$ is in fact 
unnecessarily restrictive; in a realistic physical scenario,
even in the case of locally large density variations,
the fluid accelerates with acceleration $D U / D t = a < g,$
and the relevant time scale would in fact be $t_\mathrm{ref} = U_\mathrm{ref} / a,$
i.e., the model is valid as the fluid accelerates, 
until the Mach number of the flow is no longer small.
In other words, the assumption of small Mach number is sufficient
to guarantee validity of the model.

We note two important features of the model thus far. The first is that
in both cases the perturbational pressure, which we will now denote
as $\pi(\x,t) = p(\x,t) - p_0(r)$ satisfies $\pi / p_0 = O(M^2).$ 
Thus in all but the momentum equation (where the $1 / M^2$ scaling
requires the presence of $\nabla \pi$), we can substitute $p_0$ for
$p$.  It is this approximation that decouples the pressure from
the density in such a way as to filter acoustic waves from the solution.

The second important feature is that the momentum equation
can be retained in its original form,
\[
 \frac{\partial \rho \ubold}{\partial t} + 
\nabla \cdot (\rho \ubold \ubold) + \nabla \pi
= - (\rho-\rho_0) g \er \enskip ,
\]
 with no approximation to the
buoyancy term and no assumption on the size of the perturbational density,
as long as the actual acceleration of the flow is such that Mach number of the flow remains small.

We now consider the implications of replacing $p$ by $p_0$.
The evolution of the non-reacting low Mach number system is described by the
mass and momentum equations in combination with the 
enthalpy equation, but the system remains constrained 
by the equation of state (Eq.~[\ref{eq:eos}]), namely, $p(\rho,X_k,T)= p_0(z)$.  To complete the low Mach number 
model we re-pose the equation of state as a constraint on the velocity
field, following closely the derivation in \citet{Bell:2004} but retaining
stratification effects.

We begin by re-writing conservation of mass as an expression for the
divergence of velocity:
\begin{equation}
\nabla \cdot \ubold = -\frac{1}{\rho} \frac{D \rho}{D t} \enskip .
\label{eq:divu}
\end{equation}
Differentiating the equation of state (Eq.~[\ref{eq:eos}]) 
along particle paths, we can write 
\[
\frac{Dp}{Dt}  =
  \left. \frac{\partial p}{\partial \rho}\right|_{T,X_k}    \frac{D \rho}{Dt}
+ \left. \frac{\partial p}{\partial T   }\right|_{\rho,X_k} \frac{D T}{Dt} 
+ \left. \sum_k \frac{\partial p}{\partial X_k }\right|_{\rho,T} \frac{D X_k}{Dt} \enskip ,
\]
or
\begin{equation}
\frac{D \rho}{Dt}  = \frac{1}{p_\rho}
    \left( \frac{D p}{Dt} - p_T \frac{D T}{Dt} - \sum_k p_{X_k} \omegadot_k \right) \enskip ,
\label{eq:drdt}
\end{equation}
with $p_\rho = \left.\partial p/\partial \rho\right|_{T,X_k}$, 
$p_T = \left.\partial p/\partial T\right|_{\rho,X_k}$,  and
$p_{X_k} = \left.\partial p/\partial X_k \right|_{\rho,T}$.

We now require an expression for $D T / D t$, which can be found by
differentiating the enthalpy equation (Eq.~[\ref{enthalpy_eq}]):
\[
\rho \frac{Dh}{Dt} =
  \rho \left( 
     \left.\frac{\partial h}{\partial T}\right|_{p,X_k}
     \frac{DT}{Dt} 
+    \left.\frac{\partial h}{\partial p}\right|_{T,X_k}
     \frac{D p}{Dt} 
+    \sum_k \left.\frac{\partial h}{\partial X_k}\right|_{T,p}  \frac{D X_k}{Dt} \right)
  = \frac{D p}{Dt} + \rho H \enskip .
\]
or, gathering terms, 
\begin{equation}
\frac{DT}{Dt} = \frac{1}{\rho c_p} \left( (1 - \rho h_p) \frac{D p}{D t}
- \sum_k \rho \xi_k \omegadot_k + \rho H \right) \enskip , \label{eq:dTdt}
\end{equation}
where $c_p = \left.\partial h/\partial T\right|_{p,X_k}$ is the
specific heat at constant pressure, 
$\xi_k = \left.\partial h/\partial X_k \right|_{p,T}$,
and $h_p = \left.\partial h/\partial p\right|_{T,X_k}$ 
for convenience.
Substituting equation~(\ref{eq:dTdt}) into equation~(\ref{eq:drdt}) and the resulting expression
into equation~(\ref{eq:divu}) yields 
\begin{eqnarray*}
\nabla \cdot \ubold &=& \frac{1}{\rho p_\rho} \left(
- \frac{D p}{D t} + \frac{p_T}{\rho c_p} 
  \left( (1 - \rho h_p) \frac{D p}{D t} - \rho \sum_k \xi_k \omegadot_k + \rho H \right)  
+ \sum_k p_{X_k} \omegadot_k \right)  \\
                 &=& \frac{1}{\rho p_\rho} 
  \left( \frac{p_T}{\rho c_p}(1  - \rho h_p) - 1 \right) \frac{D p}{D t}  
 + \frac{1}{\rho p_\rho} \left(
  \frac{p_T}{\rho c_p} (\rho H - \rho \sum_k \xi_k   \omegadot_k)
                               + \sum_k p_{X_k} \omegadot_k \right) \enskip .
\end{eqnarray*}
Then, replacing $p$ by $p_0(r)$, $Dp/Dt$ becomes $\ubold \cdot
\nabla p_0$, and, recalling the definition of $H$, the divergence constraint can 
be written
\begin{equation}
\nabla \cdot \ubold + \alpha \ubold \cdot \nabla p_0 =
\frac{1}{\rho p_\rho} \left(
   \frac{p_T}{\rho c_p} \left(\nabla \cdot (\kappa \nabla T) 
  - \sum_k\rho (q_k + \xi_k) \omegadot_k \right)
 + \sum_k p_{X_k} \omegadot_k \right)  \equiv \tilde{S} \enskip \label{eq:full_divu_constraint} ,
\end{equation}
where we define
\begin{equation}
\alpha(\rho,T) \equiv - \left( \frac{(1 - \rho h_p )p_T - \rho c_p}{\rho^2
  c_p p_\rho} \right) \enskip . \label{eq:alphadef}
\end{equation}
We note that for domains sufficiently smaller than a pressure scale height 
where $\nabla p_0$ can be neglected, equation~(\ref{eq:full_divu_constraint}) reduces
exactly to the divergence constraint, equation (5) in \citet{Bell:2004}.

For the larger domains that are the target of this paper, we 
use the thermodynamic identities as outlined in Appendix~\ref{sec:alpha}, 
to write
\[ \alpha = \frac{1}{\Gamma_1 p_0} \enskip , \]
where $\Gamma_1 \equiv \left. {d (\log p)}/{d (\log \rho)} \right|_s$,
and we have substituted $p_0$ for $p$.

In the case of terrestrial atmospheres and in the absence of compositional
effects, $\Gamma_1$ is replaced by the
constant $\gamma = c_p / c_v$, and using $p = \rho R T$ with $R$ the gas constant,
this expression can be simplified to 
\[ \nabla \cdot (p_0^{1/\gamma} \ubold) = \frac{R H}{c_p p_0^{R/c_p}} \enskip , \]
the pseudo-incompressible constraint as derived by \citet{durran:1989}.

For stellar atmospheres the variation in $\Gamma_1$ can be decomposed
into two contributing factors: the background stratification and the local
perturbation to that base state.   For a nearly isentropically stratified
base state and small perturbations to the base state,
$\Gamma_1$ is close to constant, hence $\rho_0 \propto p_0^{1/\Gamma_{1_0}}$. 
Neglecting expansion effects from thermal diffusion or reactions, 
this then reduces to the traditional anelastic approximation, 
\[ \nabla \cdot (\rho_0 \ubold) = 0 \enskip . \]

For the more general case $\Gamma_{1_0} = \Gamma_1(\rho_0,T_0,p_0)$
is not constant, but
we can exploit the fact that both $p_0$ and $\Gamma_{1_0}$ are functions
only of $r$.   It is straightforward (see Appendix~\ref{sec:beta}) to show that in
this case the constraint can be written 
\begin{equation}
 \nabla \cdot (\beta_0(r) \ubold) = \beta_0 \tilde{S} \enskip ,  \label{constraint_eq}
\end{equation}
where 
\[
\beta_0(r) = \beta(0) \exp \left ({\int_0^r (\frac{p_0^\prime}{\Gamma_{1_0} p_0})\,dr^\prime} \right ) \enskip .
\]

For the types of problems amenable to the low Mach number model, the
density and temperature perturbations may be large, but even so the
variation of $\Gamma_1$ due to the perturbation is at most a few percent.
For the examples in this paper we will neglect
the local variation of $\Gamma_1;$ this assumption will be re-examined 
in subsequent work.

For the purposes of comparing the fundamental hydrodynamic behavior
of the low Mach number model relative to established compressible
and anelastic formulations, we will for the remainder of this
paper neglect the effects of variation in composition,
reactions, and thermal conduction.
Summarizing the low Mach number equation set for this specialized case and re-writing the
momentum equation as an evolution equation for velocity instead, 
we now have 
\begin{eqnarray*}
\frac{\partial \rho}{\partial t}  &=& -\nabla \cdot (\rho \ubold) \enskip , \\
\frac{\partial\ubold}{\partial t} &=& -\ubold \cdot \nabla \ubold
 -\frac{1}{\rho} \nabla\pi - \frac{(\rho - \rho_0)}{\rho} g \er
 \enskip , \\
\nabla \cdot (\beta_0 \ubold )&=& 0 \enskip .
\end{eqnarray*}
We note that this system contains three equations for three 
unknowns: density, velocity and pressure.  The equation of state was
used to derive the constraint thus to include it here would
be redundant.  When reactions and compositional effects are included
in future work, the evolution equations for species and energy 
(in the form of temperature, entropy or enthalpy) will be added to this system,
but for the simple hydrodynamical tests we present here this system is
sufficient.

\section{Numerical Methodology for the Low Mach Number Model}
\label{sec:numer}

We discretize the low Mach number equation set derived in the 
previous section using an extension of the second-order accurate
projection methodology developed for incompressible flows 
\citep{bellColellaGlaz:1989,bellColellaHowell:1991,almgrenBellSzymczak:1996,
bellMarcus:1992b,almgren-iamr},
and extended to low Mach number combustion \citep{pember-flame,DayBell:2000} and 
to small-scale reacting flow for SNe Ia \citep{Bell:2004}.  
We refer the reader to the above references for numerical examples
demonstrating the second-order accuracy of the overall methodology,
as well as for many of the details of projection methods. 
Here we present a brief overview of the numerical 
methodology as applied to this equation set.
The absence of reactions and presence of $\beta_0$ in the 
projection steps are the key differences relative to the algorithm
in \citet{Bell:2004}.  

In the projection approximation, explicit discretizations of the evolution 
equations are first used to approximate
the velocity and thermodynamic variables at the new time, then an elliptic equation
for pressure, derived from the constraint imposed on the new-time velocity, 
is solved to update the pressure and return the velocity field to the 
constraint surface.  By contrast with the traditional
anelastic approach, we have not replaced conservation of mass by 
the divergence constraint, which means we are able to evolve the
density field with a 
conservative update,
rather than invoking the equation of state to diagnose it.
Thus the variables we update in each advection step
are the velocity, density, and either temperature, enthalpy,
or entropy.  
The low Mach number constraint constrains the evolution of the thermodynamic variables to
the manifold defined by the equation of state.



\begin{figure*}

\begin{center}
\epsscale{0.6}
\plotone{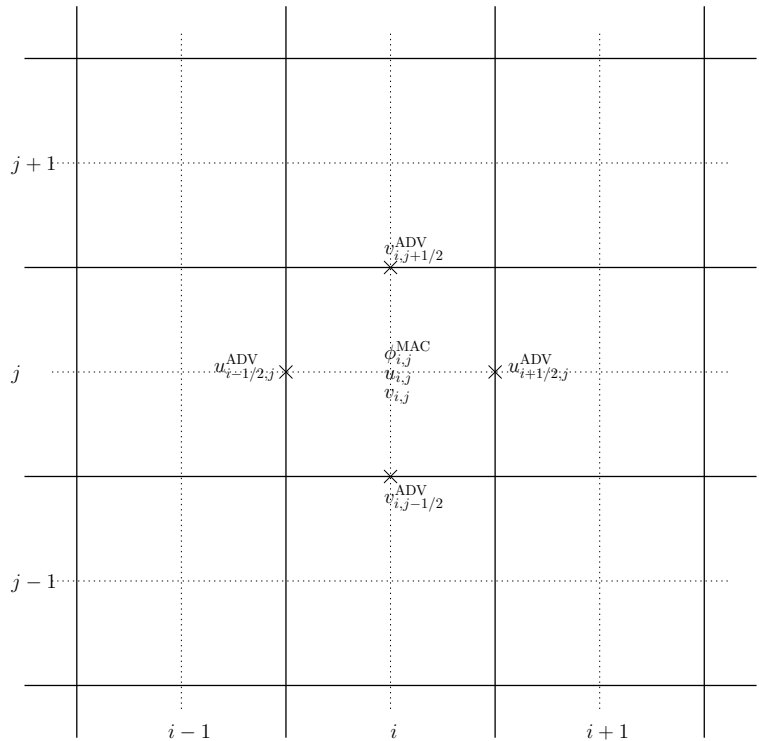}
\epsscale{1.0}
\end{center}

\caption{\label{fig:mac} Illustration of the MAC-type grid showing the
advective velocities ($u^{\mathrm{ADV}}$, $v^{\mathrm{ADV}}$) and
$\phi^{\mathrm{MAC}}$ for the $(i,j)$ cell.  An unsplit Godunov method
is used to predict these advective velocities from the cell centered velocities ($u_{i,j}$, $v_{i,j}$) in the surrounding cells.}

\end{figure*}

The discretization of the evolution equations is essentially a three-step process.
First, we use an unsplit second-order Godunov procedure \citep{colella1990} to
predict a time-centered ($t^{\nph}$) edge-based advection velocity, 
$U^{\mathrm{ADV},*}$, using the
cell-centered data at $t^n$ and the lagged pressure gradient
from the interval centered at $t^\nmh$. 
The provisional field,
$U^{\mathrm{ADV},*}$, represents a normal velocity on cell edges
analogous to a MAC-type staggered grid discretization of the
Navier-Stokes equations \citep{harlowwelch}.  Figure~\ref{fig:mac} illustrates the MAC grid.
However,
$U^{\mathrm{ADV},*}$ fails to satisfy the time-centered divergence
constraint (Eq.~[\ref{constraint_eq}]).  
We apply a discrete projection by solving the elliptic equation
\begin{equation}
 D^\mathrm{MAC} (\frac{\beta_0}{\rho^n}G^\mathrm{MAC} \phi^\mathrm{MAC}) =
D^\mathrm{MAC} (\beta_0 U^{\mathrm{ADV},*}) - \beta_0 \tilde{S}^\nph \label{eq:MAC}
\end{equation}
for $\phi^\mathrm{MAC}$, where $D^\mathrm{MAC}$ represents a centered
approximation to a cell-based divergence from edge-based velocities,
and $G^\mathrm{MAC}$ represents a centered approximation to edge-based
gradients from cell-centered data.  The solution, $\phi^\mathrm{MAC}$, is
then used to define
\[
U^\mathrm{ADV} = U^{\mathrm{ADV},*} - \frac{1}{\rho^n} G^\mathrm{MAC}
\phi^\mathrm{MAC} \enskip .
\]
$U^\mathrm{ADV}$ is a second-order accurate, staggered-grid vector field at
$t^\nph$ that discretely satisfies the constraint
(Eq.~[\ref{constraint_eq}]), and is used for computing the time-explicit
advective derivatives for $U$ and $\rho$.

We next explicitly update the density using a second-order accurate discretization
of the mass equation.  (We note here that
this approach differs from both the anelastic equation set and the alternate 
form of the low Mach number equations as described in~\citealt{Lin:2005}.)
\[ \rho^{n+1} = \rho^n - \Delta t
\left[\nabla \cdot (\rho U^\mathrm{ADV})\right]^\nph
.
\] 

The final step of the integration procedure is to advance the velocity to the new
time level.  For this step we first obtain a provisional cell-centered velocity
at $t^{n+1}$ using a time-lagged perturbational pressure gradient,
\[ \rho^\nph \frac{U^{n+1,*} - U^n}{\Delta t} + 
\rho^\nph  \left[(U^\mathrm{ADV} \cdot \nabla)
U\right]^\nph
=
- G \pi^\nmh - (\rho^\nph-\rho_0) g \er
\enskip ,
\] 
where $\rho^{\nph} = ( \rho^n + \rho^{n+1} )/2$.
At this point $U^{n+1,*}$ does not satisfy the constraint.
We apply an approximate projection to simultaneously
update the perturbational pressure and to project $U^{n+1,*}$ onto the constraint surface.
In particular, we solve
\begin{equation}
 L_\beta^\rho \phi =
   D \left( \beta_0 (\frac{U^{n+1,*}}{\Delta t} + \frac{1}{\rho^\nph} G \pi^\nmh) \right)
- \frac{\beta_0 \tilde{S}^{n+1}}{\Delta t}
\label{eq:project}
\end{equation}
for nodal values of $\phi$, where $L_\beta^\rho$ is the standard bilinear
finite element approximation to $\nabla \cdot ({\beta_0}/{\rho}) \nabla$ 
with $\rho$ evaluated at $t^\nph$. In this step, $D$ is a discrete second-order
operator that approximates the divergence at nodes from cell-centered
data, and $G = -D^T$ approximates a cell-centered gradient from
nodal data. (See \citet{almgrenBellSzymczak:1996} for a detailed discussion of this
approximate projection; see \citet{almgren:bell:crutchfield} for a discussion
of this particular form of the projection operand.)
Finally, we determine the new-time
cell-centered velocity field from
\[
U^{n+1} = U^{n+1,*} - \frac{\Delta t}{\rho^\nph}
          \left( G \phi - G\pi^\nmh \right) \enskip ,
\]
and the new time-centered perturbational pressure from
\[
  \pi^\nph = \phi \enskip .
\]
Specification of the initial value problem includes initial values for
$U$ and $\rho$ at time $t=0$ and a description of the boundary
conditions,  but the perturbational pressure is not initially prescribed. 
To begin the calculation, then, the initial velocity field is first
projected to ensure that it satisfies the divergence constraint at $t=0$.
Then initial iterations (typically two are sufficient) are performed to 
calculate an approximation to the perturbational pressure at $t = {\Delta t}/{2}$.  

In each step of the iteration we follow the procedure described above.
In the first iteration we use $\pi^{-\myhalf} = 0$. 
At the end of each iteration we have calculated a new value of $U^1$ and
a pressure $\pi^{\myhalf}$.  During the iteration procedure, we discard
the value of $U^1$, but define $\pi^{-\myhalf} = \pi^{\myhalf}$.  Once the
iteration is completed, the above algorithm can be followed as written.

\section{Numerical Validation and Comparison}
\label{sec:compare}

\subsection{Compressible Formulations}

We compare and contrast the low Mach number results with those
obtained using two different discretizations of the fully compressible
equation set, both implemented in the FLASH Code~\citep{flash}. 
The first is the piecewise parabolic method
(PPM)~\citep{ppm}, which is a high-order accurate, dimensionally split algorithm
where the updates are done in one-dimensional sweeps, e.g. in two
dimensions
\[
S_{i,j}^{n+2} = X(\Delta t) Y(\Delta t) Y(\Delta t) X(\Delta t)
S_{i,j}^n \enskip ,
\]
where $S = (\rho, \rho \ubold, \rho E)$ is the state variable,
$X(\Delta t)$ is the operator that updates the state through
$\Delta t$ in time in the horizontal direction, and 
$Y(\Delta t)$ updates the state by $\Delta t$ in the vertical direction.
One PPM cycle updates the
state through $2\Delta t$, switching the order of the directional
operators midway through to retain second-order accuracy.  PPM is the
primary hydrodynamics algorithm used by the large-scale SNe~Ia explosion
modeling community~\citep{roepke2005,plewa:2004,gamezo:2003}.
The FLASH implementation of PPM has been well validated \citep{flash-validation},
and serves as a good basis for comparisons with the low Mach number
algorithm.

A numerical issue that arises in fully compressible simulations,
but not with the low Mach number approach, is the difficulty of maintaining
a quiet hydrostatic atmosphere.  Small displacements from 
hydrostatic equilibrium (HSE) can generate sound waves 
throughout the atmosphere, which, if unchecked, can lead to ambient velocities 
that can swamp the process being studied.  
The hydrostatic equilibrium improvements described in~\citet{ppm-hse},
which remove the hydrostatic pressure from the pressure jump
across the interfaces in the Riemann problem,  
were used for all PPM runs.  
The upper and lower boundary conditions are
the hydrostatic boundaries described in that same paper, with
the pressure and density modified according to hydrostatic
equilibrium, and the velocities given a zero gradient.  


The second compressible algorithm we consider is a second-order unsplit method
following \citet{colella1990}.
At low Mach number, dimensionally split methods can have trouble producing 
realistic velocity fields,  as will be shown in the bubble rise
comparison.  
In the unsplit formulation, the cell averages are updated in all
directions at once.  A critical part of the unsplit method is that the
interface reconstructions contain a transverse flux term that
explicitly couples in the information from the corner cells.  
Fourth-order accurate slope limiting is used in the central 
differences, as well as high-order reconstruction of the states for 
the transverse Riemann problem, as described in \citep{colella1985}.
This is the same procedure used in predicting the interface states in the
low Mach number method presented here. 
This method was extended to handle general equations of
state following the procedure in~\citet{colellaglaz1985}, adding an
additional transverse flux piece to the interface reconstruction of
$\gamma$, to be consistent with the unsplit reconstructions.  This was put into the FLASH framework for the present simulations.  Both the
PPM and unsplit solvers use the same two-shock Riemann solver
described in~\citet{colellaglaz1985}.
For both the split and unsplit solvers, a CFL number of~0.8 was used
based on the sound speed.

\subsection{Anelastic Approach}

The low Mach number equations and the anelastic equation set are
derived differently.  Both equation sets assume a low Mach number,
equivalently a small pressure perturbation from the background state.  
However,  the anelastic equation set assumes 
both small density and small temperature perturbations as well.
As noted earlier the velocity constraints resulting from these two derivations 
are strikingly similar, and in fact equivalent for an isentropically stratified 
background state.  Even in the non-isentropic
background considered here, the differences between $\rho_0$ and
$\beta_0$ are small. 

However, because the anelastic approximation assumes small density and 
temperature perturbations, approximations are made to the buoyancy
term in the momentum equation.  These follow from
the observation that since the perturbational density was neglected
in the continuity equation in order to derive the velocity constraint,
the continuity equation cannot be used to evolve the perturbational density.
Thus an alternative formulation of the buoyancy term must be used.
A typical anelastic model evolves temperature or entropy, and constructs the buoyancy
forcing term from that field using a linearized approximation.
Following the derivation by \citet{BraRob:1995} that combines parts of the
pressure gradient and buoyancy terms, we consider the following
form of the anelastic equations:
%
\begin{eqnarray*}
 \rho_0 \frac{D \ubold}{D t} &=& - \rho_0 \nabla \left (\frac{p^\prime}{\rho_0} \right)
- \left( \frac{\partial \rho_0}{\partial S}\right)_p S^\prime g \er \enskip , \\
\frac{D S}{D t} &=& 0 \enskip , \\
\nabla \cdot (\rho_0 \ubold )&=& 0 \enskip .
\end{eqnarray*}
Here $S$ is entropy, $S^\prime = S - S_0$, $p^\prime = p - p_0$,
and we have neglected viscosity, thermal diffusivity and the gravitational 
potential perturbation.   
We note that in the case of small density and temperature perturbations,
simulations using the low Mach number equations and the anelastic approximation
give indistinguishable results, thus for the numerical comparisons we focus on problems with
finite amplitude perturbations as described in the next subsection. 

 

\subsection{Bubble Rise Comparison}

\begin{figure*}

\begin{center}
\plotone{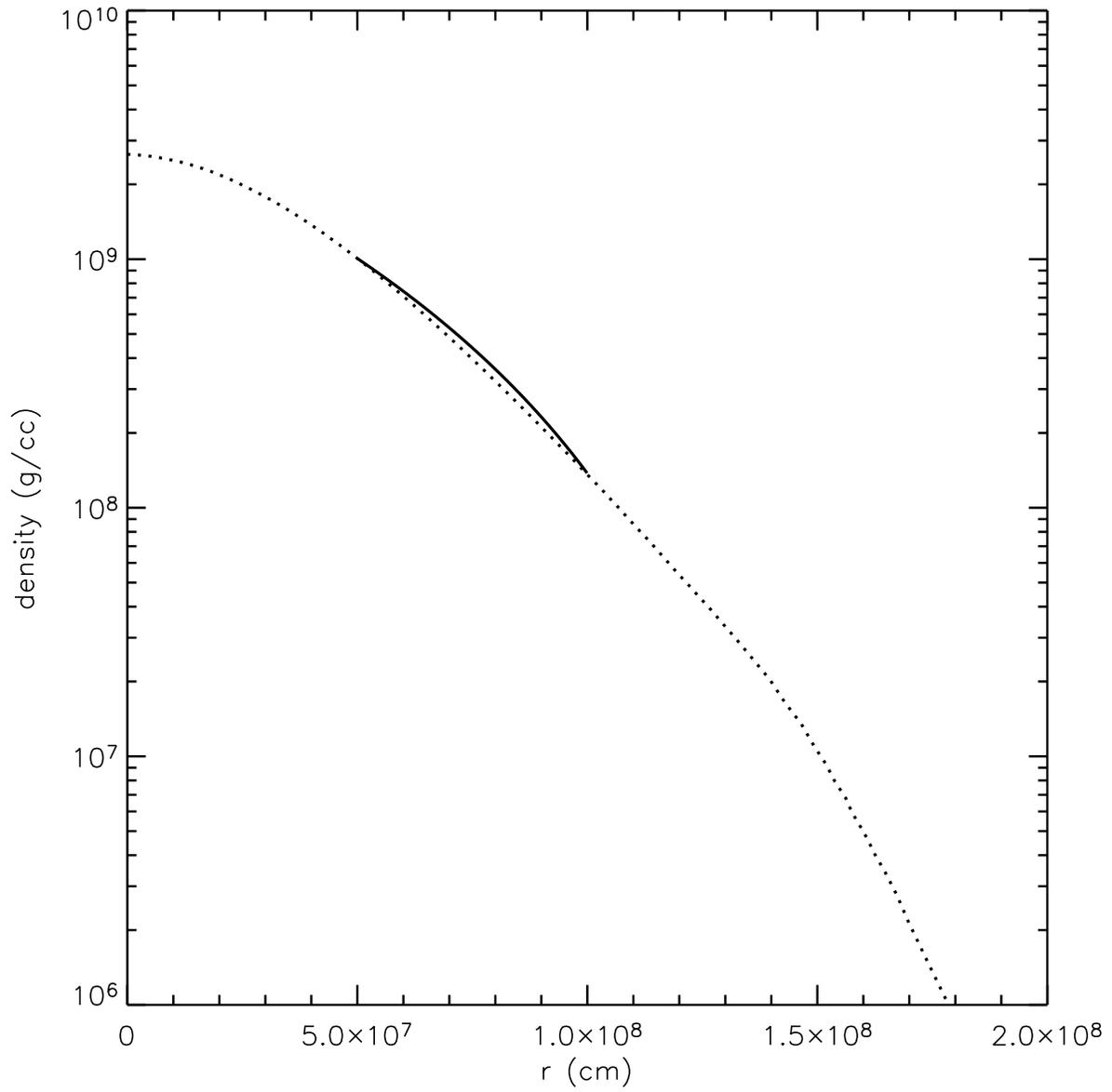}
\end{center}

\caption{Initial model generated by the Kepler code
(dotted) and uniformly gridded, constant gravity portion used in the
bubble simulations (solid).}
\label{fig:model} 
\end{figure*}

We present three sets of two-dimensional calculations
of a rising bubble in a stellar environment.  The one-dimensional 
background state ($\rho_\mathrm{0}$, $T_\mathrm{0}$, $p_\mathrm{0}$) is calculated 
using the Kepler code~\citep{weaver:1978}, to evolve a Chandrasekhar mass white
dwarf until the central temperature reaches $7\times 10^8$~K, 
representing conditions just before ignition.  We map a portion of the one-dimensional model onto
a uniform two-dimensional grid, and place it into hydrostatic equilibrium with a
constant gravitational acceleration ($g = -1.9\times
10^{10}~\mathrm{cm}~\mathrm{s}^{-2}$).
We further simplify by ignoring metric terms associated with the radial coordinate and view
the domain as Cartesian.
We note that neither the constant gravity assumption nor the simplified metric is
a limitation of any of the methods presented here, but is chosen
in these comparison simulations for simplicity.  The density structure of the model is illustrated in
Figure~\ref{fig:model}.

All bubbles begin in pressure equilibrium with the
background state and are defined by a simple temperature perturbation,
from which the density perturbation is calculated.
We consider three different cases, which we will distinguish by the
maximum temperature at the center of the bubble, $T_\mathrm{max}$.
The temperature profile of the bubble is then defined by
\[
T = T_\mathrm{0} +  (T_\mathrm{max}-T_\mathrm{0}) \frac{1}{2} \left( 1 + \tanh\left( 
\frac{2.0-{\xi}/{\delta}}{0.9} \right) \right) \enskip ,
\]
where 
\[
\xi = \sqrt{(x-x_\mathrm{cent})^2 + (r-r_\mathrm{cent})^2} \enskip ,
\]
and
$(x_\mathrm{cent},r_\mathrm{cent}) = (2.5\times10^7,6.25\times
10^7)$~cm, $\delta = 1.25\times 10^6$~cm in a domain 
from $x=0$~cm to $5\times 10^7$~cm and $r=5\times 10^7$~cm to $10^8$~cm. 
The stellar equation of state is then used to
compute $\rho$ given $T$ and $p_\mathrm{0}$. This profile was
chosen to give a smooth transition from the ambient temperature to the
perturbed temperature, thus minimizing the effects of the numerical slope 
limiters present in the different hydrodynamics methods.  Due to the short
time scale of the problem thermal diffusivity is neglected. For all bubble
calculations presented the grid has uniform resolution of 384 x 384; the 
adaptive gridding features of all the codes are turned off.

\begin{figure*}
\begin{center}
\epsscale{0.85}
\plotone{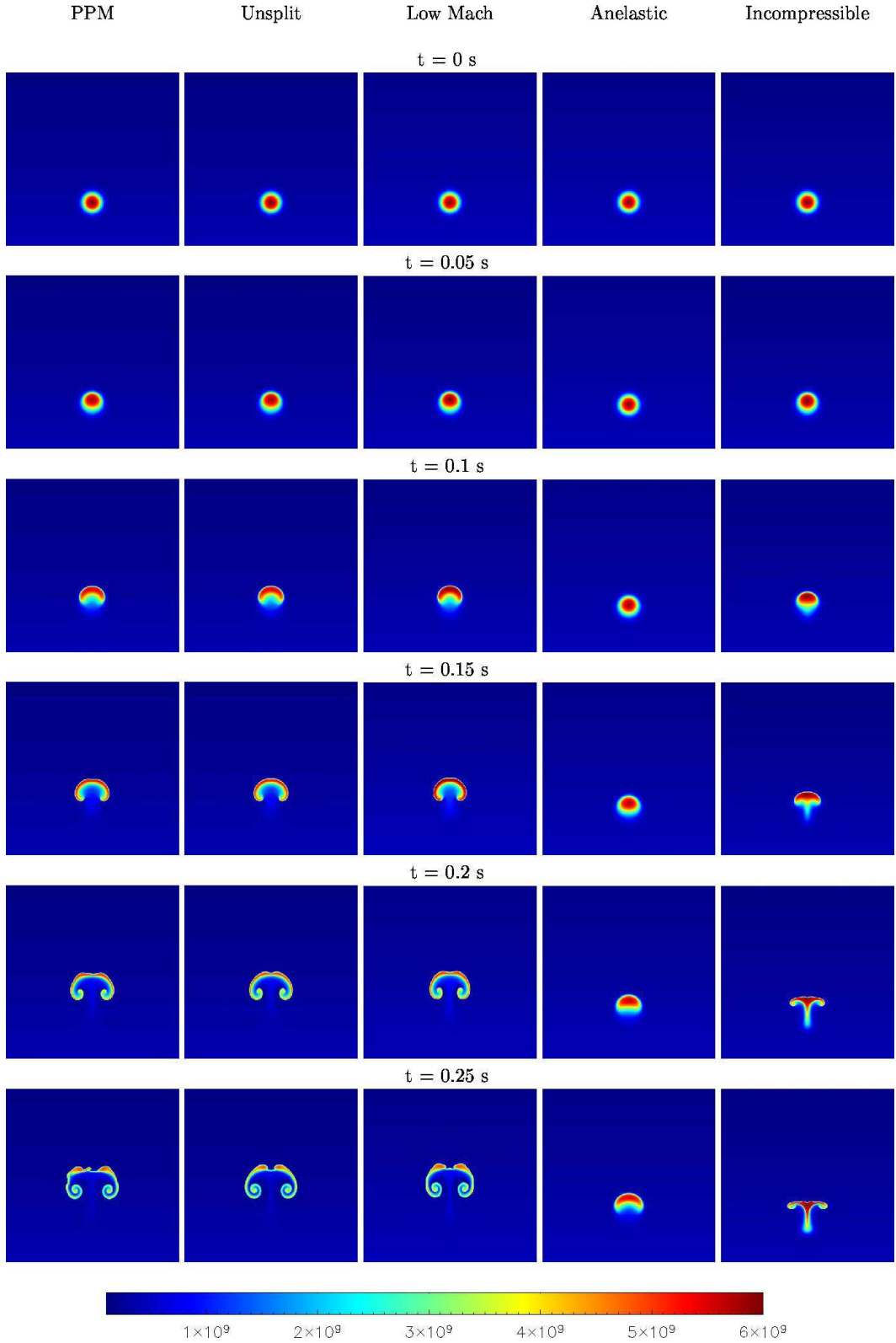}
\epsscale{1.0}
\end{center}

\caption{Bubble evolution for five different
algorithms.  Here, the peak temperature is $6\times 10^9$~K.  We see good agreement between the two compressible codes (PPM and unsplit) and the low Mach number algorithm.}
\label{fig:bubble_6.e9} 
\end{figure*}

\begin{figure*}
\begin{center}
\epsscale{0.8}
\plotone{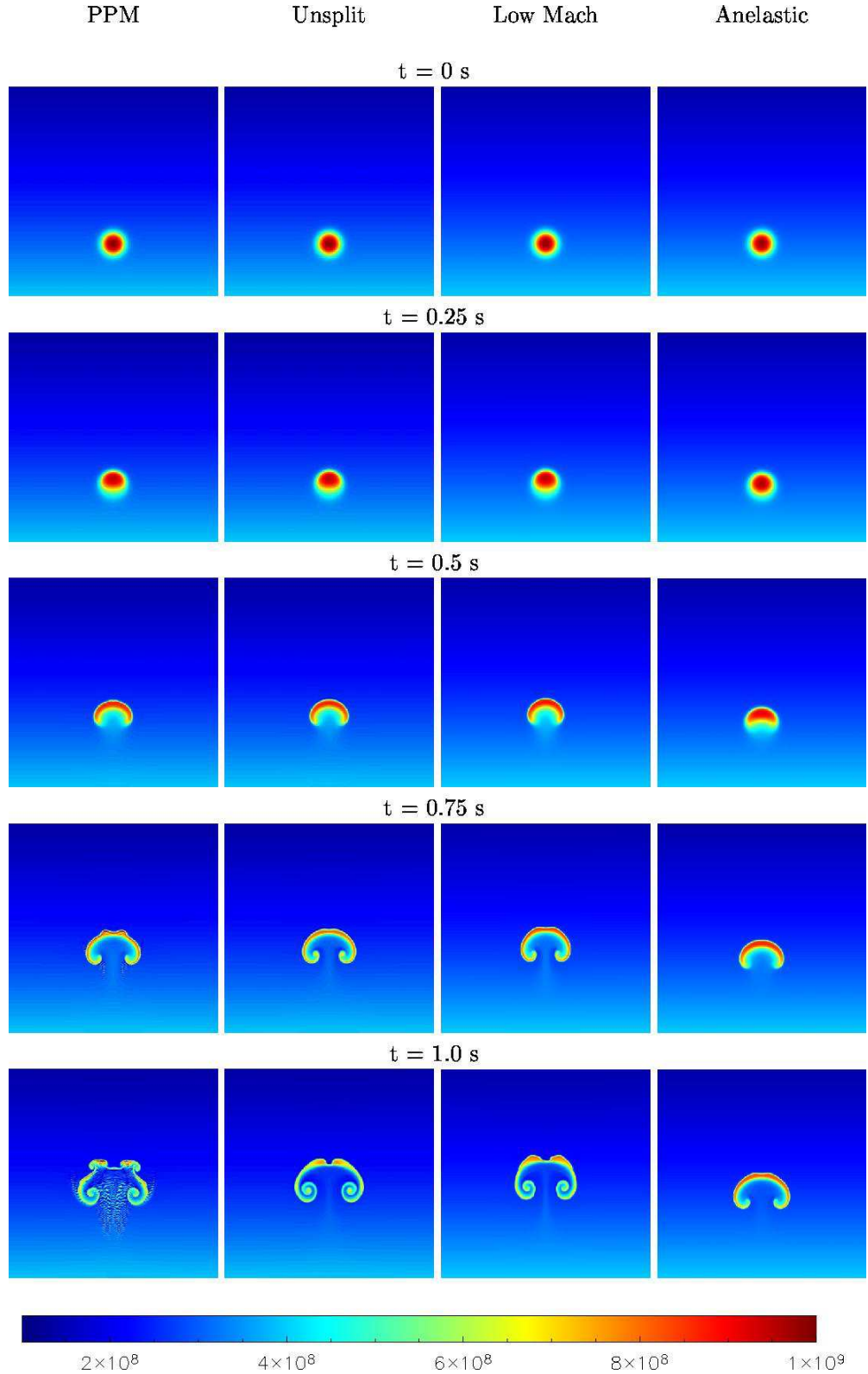}
\epsscale{1.0}
\end{center}

\caption{Bubble evolution for four different
algorithms.  Here, the peak temperature is $1\times 10^9$~K.  As with the
$6\times 10^9$~K case, the PPM, unsplit, and low Mach number results are 
in good agreement.}
\label{fig:bubble_1.e9} 
\end{figure*}

Figures \ref{fig:bubble_6.e9} and \ref{fig:bubble_1.e9}
present comparisons of simulations using the
low Mach number approach, two different discretizations of the fully 
compressible equation set, as well as the anelastic and incompressible
equation sets.  The low Mach number, anelastic and incompressible results 
are calculated using the projection method approach described in the 
previous section. The only differences in methodology occur in the
coefficient of velocity in the projection, and in the construction of the
buoyancy term.  Each of these methods was run at a CFL number of~0.9 
based on the maximum advection velocity.

Figure~\ref{fig:bubble_6.e9} shows the temperature evolution for 
$T_\mathrm{max} = 6\times 10^9$~K.  This
corresponds to an Atwood number for the bubble of approximately 0.079.
In this simulation, the bubble reaches a Mach number of about~0.2.  In
addition to the PPM and unsplit compressible solvers, traditional
anelastic and incompressible solvers are shown for comparison.  
The low Mach number method closely tracks the
two compressible solvers.  The incompressible and anelastic results
demonstrate the effects of their respective assumptions.  The 
velocity constraint for the anelastic model is sufficiently accurate
to capture the bubble rise, but because of the linearity
of the density-temperature relationship in the anelastic approach,
the buoyancy term is too small in the anelastic simulation.
By contrast, the incompressible simulation contains the full buoyancy 
term but due to the incompressibility constraint the bubble cannot expand; 
consequently, it reaches neutral buoyancy at a much lower level and stops rising.
We do not follow the bubbles past the point where nonlinear instabilities 
along the sides begin to dominate the evolution.

\begin{figure*}
\begin{center}
\epsscale{1.0}
\plotone{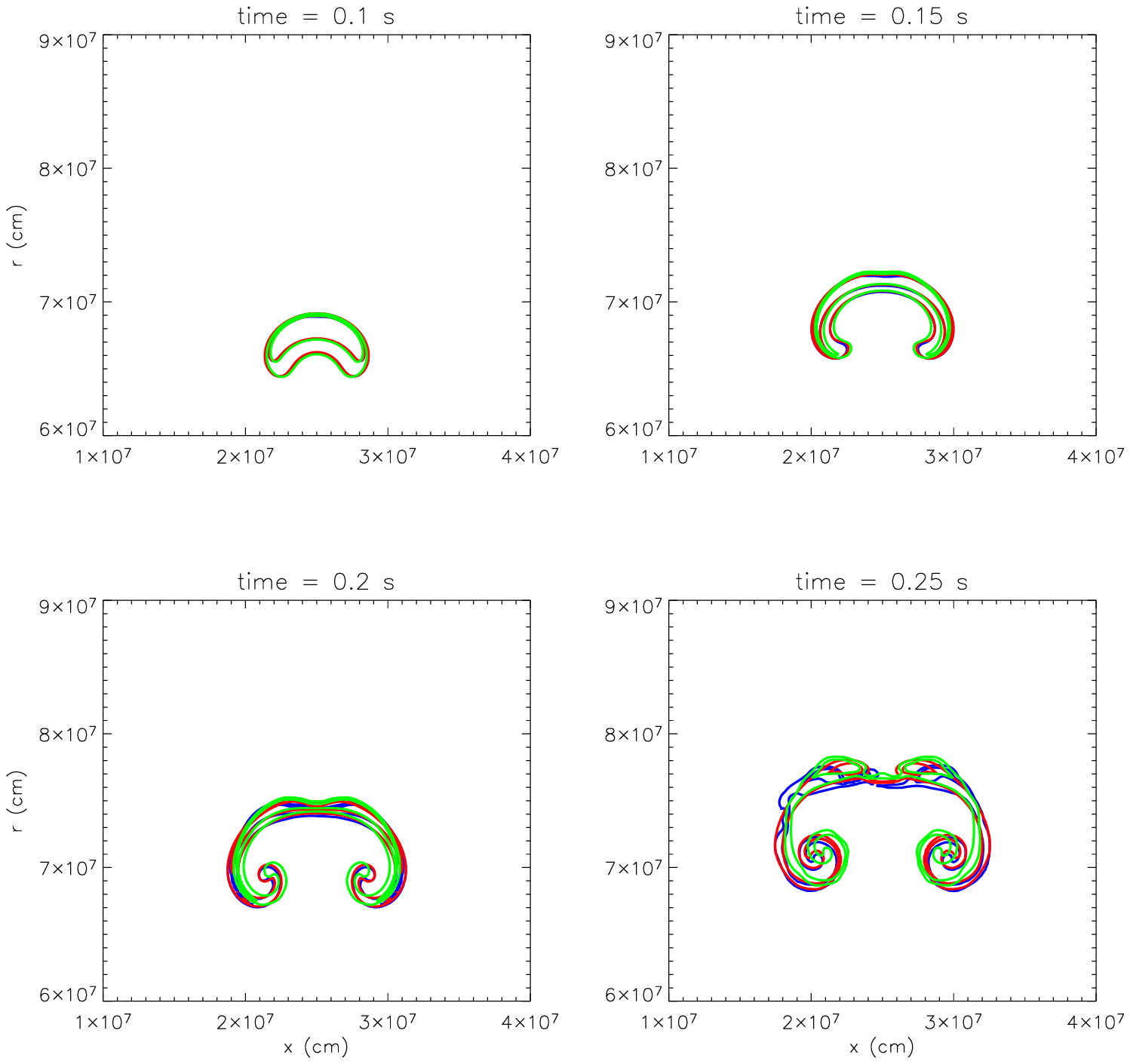}
\epsscale{1.0}
\end{center}

\caption{Detailed comparison of the
temperature field for PPM (blue), the unsplit compressible algorithm (red), and the
low Mach number algorithm (green), for the $6\times 10^9$~K
perturbation.  Contours are shown at $3\times 10^9$ and $4.5\times 10^9$~K. }
\label{fig:compare_6.e9} 
\end{figure*}

\begin{figure*}
\begin{center}
\plotone{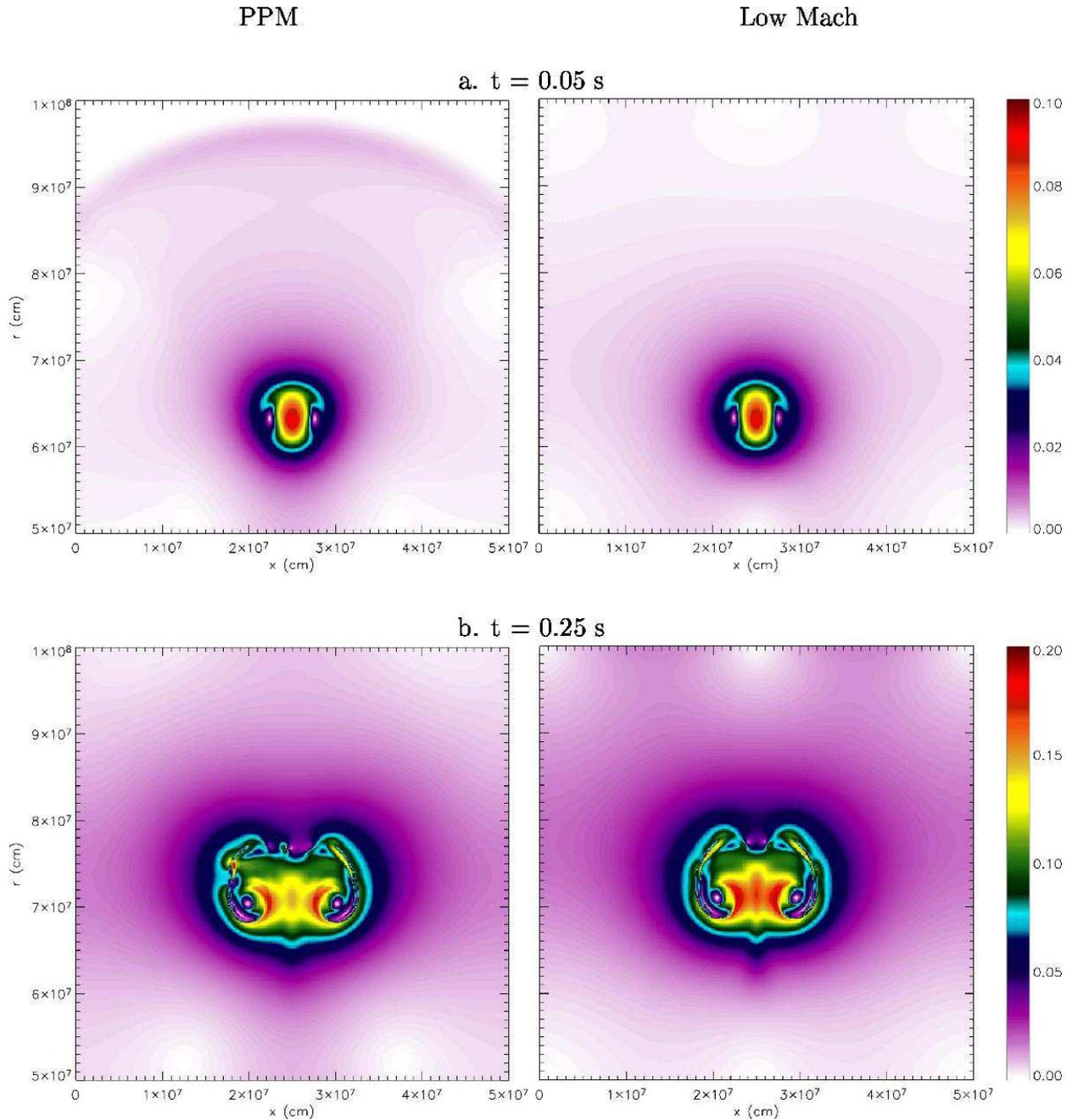}
\end{center}
\caption{Mach number comparison for the $6\times 10^9$~K bubble.  At
early times (a.), the PPM results show a sound wave from the initial
perturbation just about to exit through the top of the domain.  This
is not present in the low Mach number case, as it filters sound waves.
At late times (b.), the flow has attained a Mach number of almost 0.2
in places.}
\label{fig:mach}
\end{figure*}

A more detailed comparison of the results from Figure~\ref{fig:bubble_6.e9} 
is provided in Figure~\ref{fig:compare_6.e9},
where temperature contours of the low Mach number solution are
superimposed on temperature contours of the the unsplit and PPM 
solutions.  Here we see a large degree of overlap, demonstrating that 
the bubbles have the same rise velocity and size independent of the 
algorithm.  Figure~\ref{fig:mach} shows the Mach number of the PPM
and low Mach number methods, further demonstrating the agreement
between the two sets of results, with the exception of the unphysical
loss of symmetry in the PPM simulation.

Figure~\ref{fig:bubble_1.e9} shows the temperature evolution for the
lower peak temperature case, $T_\mathrm{max} = 1\times 10^9$~K,
and corresponding Atwood number of~0.0024.
Over the course of this comparison, the
Mach number remains below~0.05.  Again, we observe the agreement
between the low Mach number and compressible solvers, with the
exception of the late-time breakdown of the PPM solution,
indicated by the large amplitude temperature oscillations 
dominating the flow behind the bubble.
These oscillations reflect the poor performance of operator split
algorithms for very low speed flows.

\begin{figure*}
\begin{center}
\epsscale{1.0}
\plotone{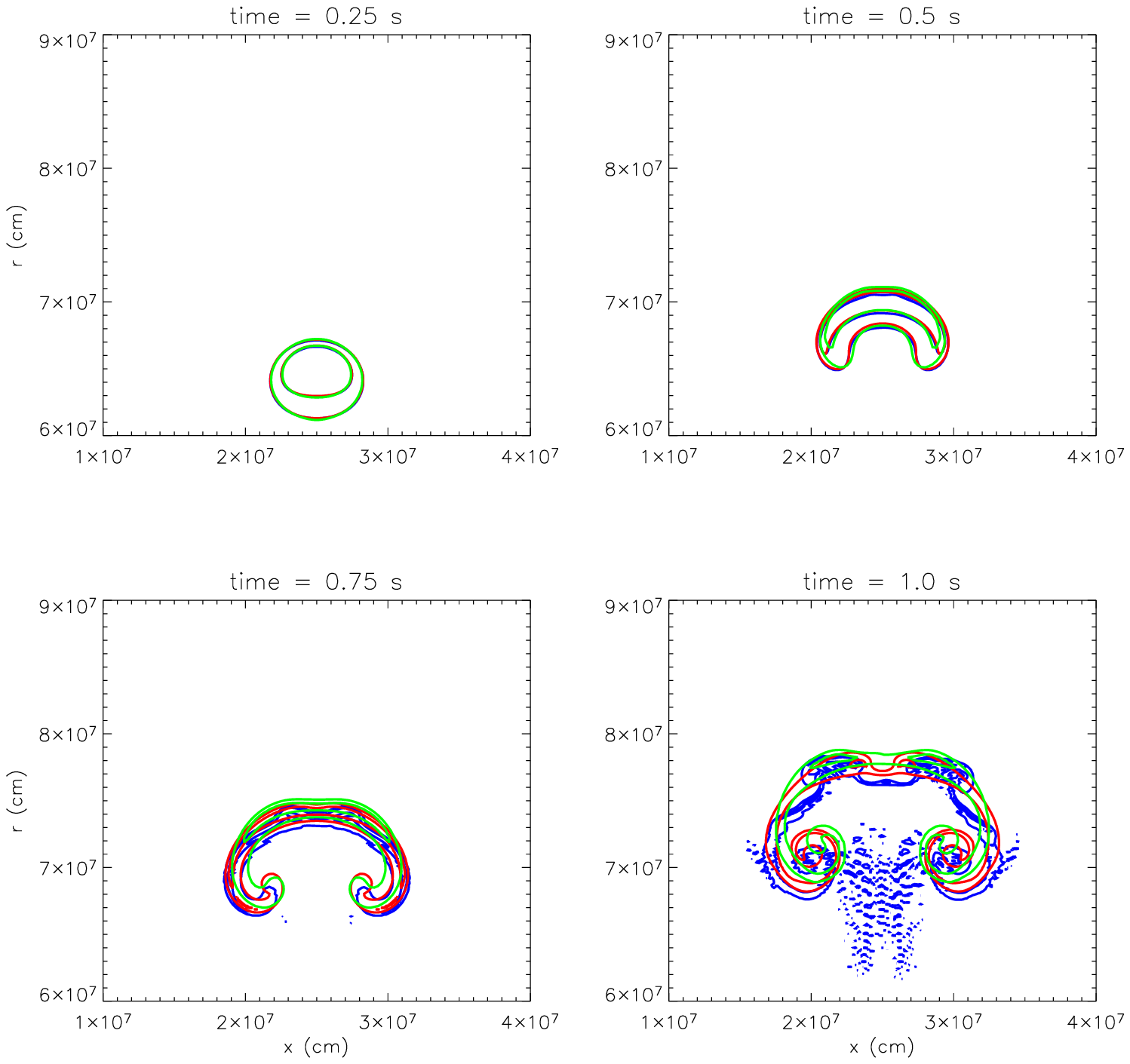}
\epsscale{1.0}
\end{center}

\caption{Detailed comparison of the
temperature field for PPM (blue), the unsplit compressible algorithm (red), and the
low Mach number algorithm (green), for the $1\times 10^9$~K
perturbation.  Contours are shown at $5\times 10^8$ and $7.5\times 10^8$~K. }
\label{fig:compare_1.e9} 
\end{figure*}

A more detailed comparison of the results from Figure~\ref{fig:bubble_1.e9} 
is shown in Figure~\ref{fig:compare_1.e9}, again superimposing 
temperature contours from the low Mach number and compressible
formulations.  We again see good agreement, with the exception of the breakdown of PPM at late times.

Timings of the PPM and low Mach number code were made on a single
processor (1.53~GHz Athlon MP) using the Intel 8.1 compilers.  Both
codes were compiled with the same compiler optimization flags: \texttt{-O3 -ip -ipo}.  FLASH was set to run with $16\times 16$ zone blocks, instead of the default $8\times 8$ for better performance with uniform gridding.
The $6\times 10^9$~K bubble required 2148 timesteps, taking 14200~s to
evolve the bubble to $0.25$~s in simulation time.  By comparison, the
low Mach number solver took 246 timesteps and 1480~s, about an order
of magnitude speed-up.  For the
$10^9$~K bubble, the PPM solver took 7842 timesteps, taking 52100~s to
evolve the bubble for 1~s of simulation time, while the low Mach
number solver took 252 timesteps and 1560~s.  As expected, the
performance gap increases as the Mach number decreases.  The unsplit compressible algorithm takes approximately twice as much time to run as PPM, primarily due to the additional transverse Riemann solves required.

\begin{figure*}
\begin{center}
\epsscale{0.5}
\plotone{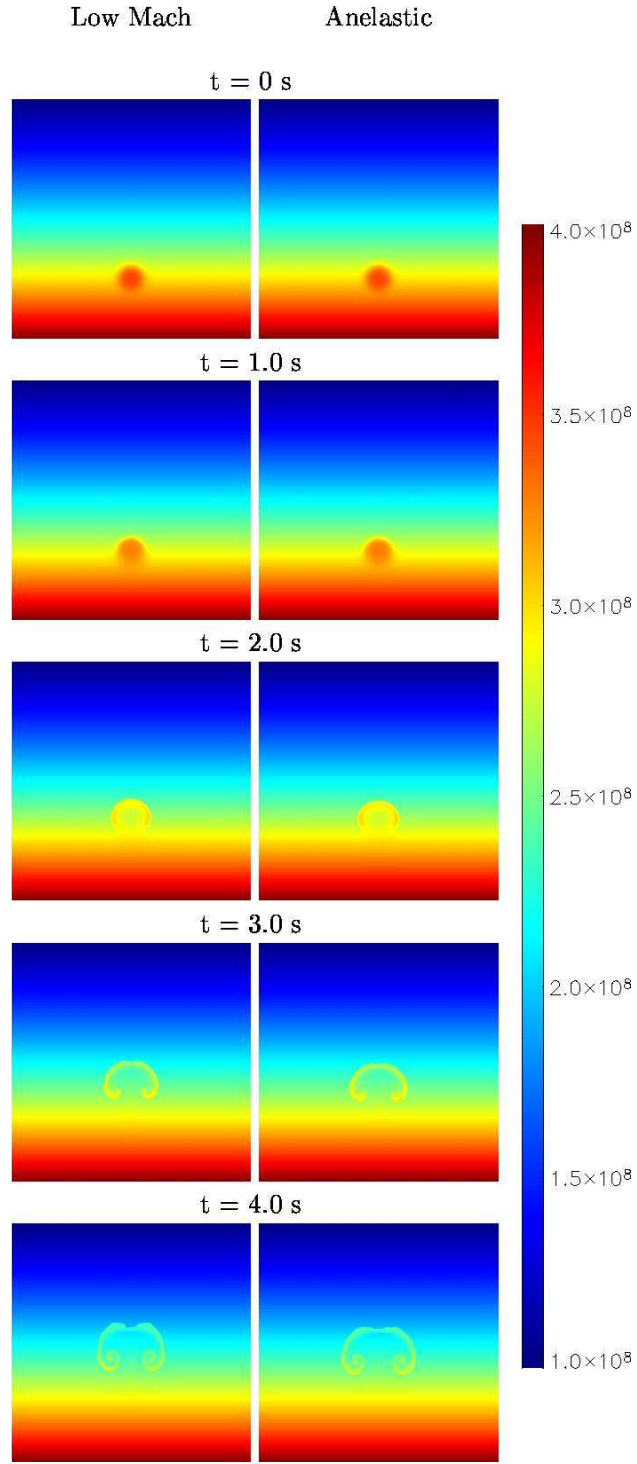}
\epsscale{1.0}
\end{center}

\caption{Bubble evolution the low Mach number
and anelastic algorithms, showing good agreement for the two different methods..  Here, the peak temperature is $3.5\times
10^8$~K.  }
\label{fig:bubble_3.5e8} 
\end{figure*}

\begin{figure*}
\begin{center}
\epsscale{1.0}
\plotone{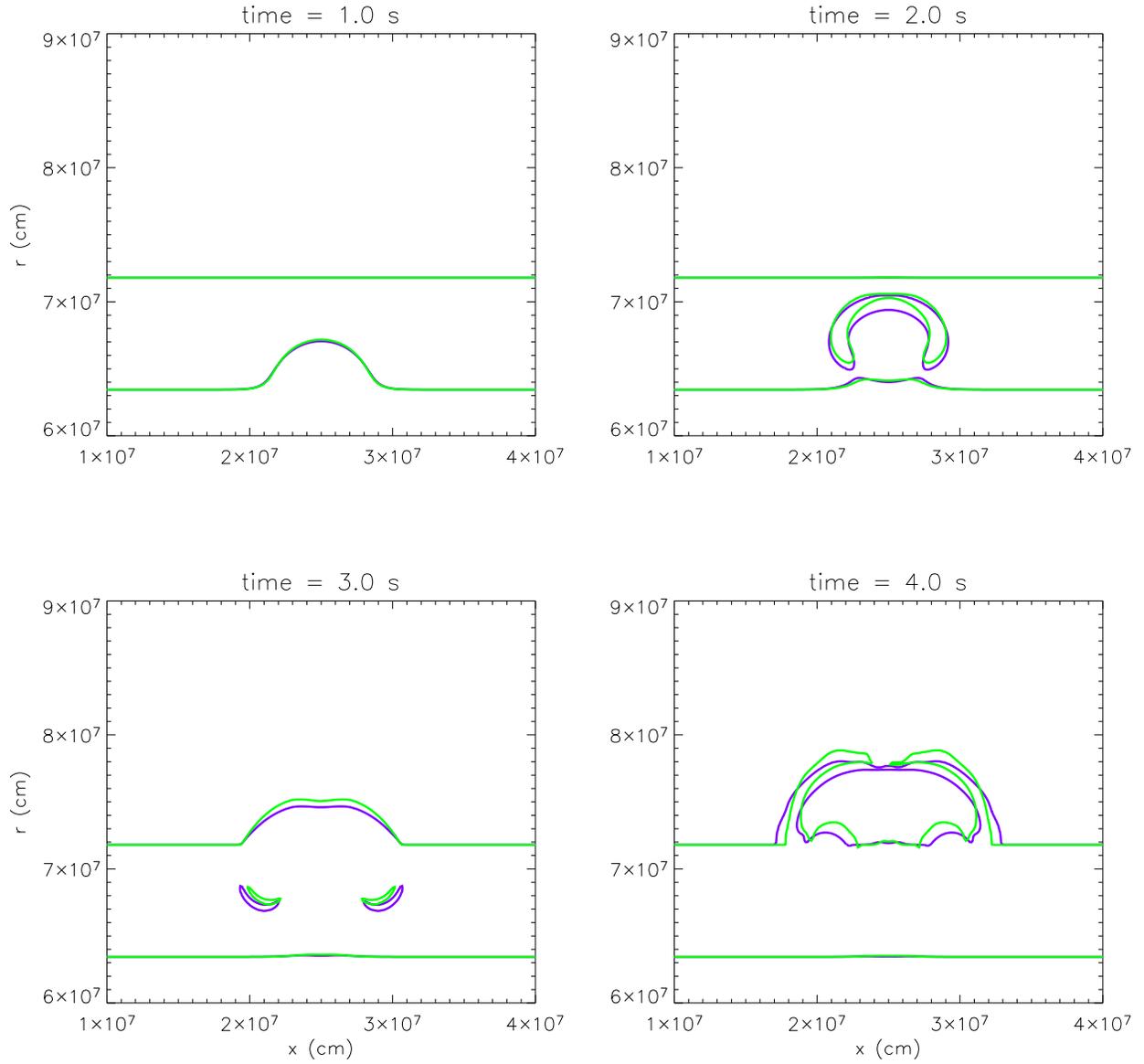}
\epsscale{1.0}
\end{center}

\caption{Detailed comparison of the
temperature field for the anelastic algorithm (purple) and the
low Mach number algorithm (green), for the $3.5\times 10^8$~K
perturbation.  The contours are at $2.5\times 10^8$ and $3\times 10^8$~K}
\label{fig:compare_unsplit_3.5e8} 
\end{figure*}

Finally, in Figures \ref{fig:bubble_3.5e8} and \ref{fig:compare_unsplit_3.5e8}
we compare the low Mach number
and anelastic models for a bubble with $T_\mathrm{max} = 3.5\times 10^8$~K.
This regime is inaccessible to the compressible formulation, with a peak
Mach number during the calculation of $0.012$.
We note that, as expected, as the Atwood number decreases the
fidelity of the anelastic approximation improves. 

The three cases presented in this section demonstrate successful
application of the low Mach number approach, as well as 
failure of the compressible approach for low-speed flows,
and failure of the anelastic approach for flows with large density or temperature
variations.  The low Mach number approach, like
the other methods, has limits to its applicability. Specifically,
as the flow speed, hence the Mach number, increases, typically the
thermodynamic pressure will diverge from $p_0$ and the assumptions
underlying the low Mach number approach will be violated. 
As with the anelastic model, numerical simulations using the low Mach number
model will continue to yield what appear to be reasonable 
solutions even as the underlying assumptions are violated,
but the solutions will no longer be physically relevant.  

In the case of a bubble rise without heat sources,
it is difficult to numerically demonstrate the failure of the low
Mach number method even as the Mach number increases.  
However, the divergence of the low Mach number model from the compressible
solution in the case of large Mach number will be discussed more
thoroughly in the subsequent paper that discusses the behavior
of the low Mach number model in the presence of heating.


\section{Conclusions}

We have introduced a new method for following low speed, stratified flows in 
astrophysical conditions and 
have demonstrated, through comparison with compressible and anelastic 
codes, that this algorithm performs well in the range of Mach number
from near zero to about 0.2.
The increased computational efficiency associated with a 
low Mach number formulation makes it an ideal 
tool for investigations of the
convective/ignition phase of SNe Ia.  
However, to be applicable in this setting, a number of generalizations to 
the methodology will need to be developed. In particular, 
we will need to extend the the
algorithm to include the effects of variation in composition,
reactions and thermal conduction.
In addition, once a flame is established it will be necessary to include sub-grid models 
for turbulent flame propagation that will enable the methodology to be used 
to simulate the evolution of the early 
phases of the explosion.  
These issues will be addressed in future work.

\acknowledgments

We thank Gary Glatzmaier, Michael Kuhlen, and Tami Rogers for helpful
discussions regarding the anelastic hydrodynamics method, and Alan
Calder and Jonathan Dursi for helpful comments on the manuscript.  We
especially thank Stan Woosley for numerous discussions and
interactions.  The PPM calculations presented here used the FLASH Code
(version 2.5), developed in part by the DOE-supported ASC/Alliance
Center for Astrophysical Thermonuclear Flashes at the University of
Chicago.  This work was supported by the Applied Mathematics Program
of the DOE Office of Mathematics, Information, and Computational
Sciences under the U.S. Department of Energy under contract No.\
DE-AC03-76SF00098, by DOE grant No.\ DE-FC02-01ER41176 to the
Supernova Science Center/UCSC, and by the NASA Theory Program 
(NAGW-5-12036).

\appendix
\section{Simplification of $\alpha$}
\label{sec:alpha}

In this appendix we derive a simplified expression for
$\alpha$ introduced in equation~\ref{eq:alphadef}.
We refer to Chapter 9 of~\cite{cox-giuli-v1} (CG, in
this appendix),
for a number of thermodynamic identities.

We begin by rewriting $h_p$, using
\[
\frac{\partial h}{\partial \rho} \Bigg |_T = 
\frac{\partial h}{\partial p} \Bigg |_T 
\frac{\partial p}{\partial \rho} 
= h_p p_\rho \enskip .
\]
with
\[
h(\rho, T) = \frac{p(\rho, T)}{\rho} + e(\rho, T) \enskip .
\]
Therefore, 
\[
h_p = p_\rho^{-1}\left \{ {-\frac{p}{\rho^2} + \frac{p_\rho}{\rho} + e_\rho} \right \} =
\frac{1}{\rho} \left ( 1 - \frac{p}{\rho p_\rho} \right ) + \frac{e_\rho}{p_\rho} \enskip .
\]
Putting this into $\alpha$, we have
\begin{eqnarray*}
\alpha &=& - \frac{1}{\rho^2 c_p p_\rho} \left 
[ \left ( 1 - \left (1 - \frac{p}{\rho p_\rho} \right ) - \rho \frac{e_\rho}{p_\rho} \right ) p_T - \rho c_p \right ] \\
 &=& - \frac{1}{\rho p_\rho c_p} \left [ \left ( \frac{p}{\rho^2 p_\rho} - \frac{e_\rho}{p_\rho} \right ) p_T - c_p \right ] \enskip .
\end{eqnarray*} 

For a generalized equation of state, there are three principal
adiabatic exponents which relate the various differentials (d$p$,
d$T$, and d$\rho$).  For an ideal gas, they are all equivalent.  Here,
we use $\Gamma_1$ (CG~eq.~9.88):
\[
\Gamma_1 \equiv \left ( \frac{d \ln p}{d \ln \rho} \right )_{\mathrm{ad}} \enskip .
\]
This is related to the ratio of specific heats, $\gamma$ via
\[
\gamma = \frac{c_p}{c_V} = \frac{\Gamma_1}{\chi_\rho} \enskip ,
\]
(CG~eq.~9.87) where
\[
\chi_\rho \equiv \left ( \frac{\partial \ln p}{\partial \ln \rho} \right )_T 
= \frac{\rho}{p} p_\rho
\]
(CG~eq.~9.82) is the ``density exponent in the pressure equation of
state.''
For an ideal gas, $\chi_\rho = 1$, and $\Gamma_1$ = $\gamma$.
Taking all of this together, we see that
\[
\frac{1}{\rho p_\rho} = \frac{\gamma}{\Gamma_1 p} \enskip .
\]
Putting this into our $\alpha$ expression,
\begin{equation}
\alpha = - \frac{\gamma}{\Gamma_1 p c_p} 
\left [ 
  \left ( \frac{p}{\rho^2 p_\rho} - \frac{e_\rho}{p_\rho} \right ) 
  p_T - c_p
\right ] \enskip .
\label{eq:alpha}
\end{equation}
Motivated by the ideal gas result that $\alpha = 1/(\gamma p)$, we
want to show that the quantity in the square brackets
in equation~(\ref{eq:alpha})  reduces to $c_V$.

The specific heats are related by (CG~eq.~9.84)
\begin{equation}
c_p - c_V = -\frac{E}{T} \left (\frac{\partial \ln E}{\partial \ln \rho} \right )_T \frac{\chi_T}{\chi_\rho} + \frac{p}{\rho T} \frac{\chi_T}{\chi_\rho} \enskip ,
\label{eq:heats}
\end{equation}
The temperature exponent is defined as
\[
\chi_T \equiv \left ( \frac{\partial \ln p}{\partial \ln T} \right )_\rho
= \frac{T}{p} p_T \enskip
\]
(CG~eq.~9.81), so equation~(\ref{eq:heats}) simplifies to
\begin{eqnarray*}
c_p - c_V &=& - \frac{\rho}{T} e_\rho \frac{\chi_T}{\chi_\rho} +
\frac{p}{\rho T} \frac{\chi_T}{\chi_\rho} \\
&=& - e_\rho \frac{p_T}{p_\rho} + \frac{p}{\rho^2} \frac{p_T}{p_\rho} \enskip ,
\end{eqnarray*}
which substitutes directly into equation~(\ref{eq:alpha}) to yield
\[
\alpha = - \frac{\gamma}{\Gamma_1 p c_p} \left [ (c_p - c_V) - c_p \right ]
= - \frac{\gamma}{\Gamma_1 p c_p} \left [ - c_V \right ] 
= \frac{1}{\Gamma_1 p} \enskip .
\]
We note that $\Gamma_1$ varies slowly throughout the white dwarf and
is a quantity that is already returned by the tabular equation of
state.

\section{Derivation of $\beta$}
\label{sec:beta}

We seek a function $\beta(z)$ such that
\[
\frac{1}{\beta(z)} \nabla \cdot ( \beta \ubold )
 = (\nabla \cdot \ubold) + \frac{1}{\Gamma_1 p_0} \ubold \cdot \nabla p_0 \enskip .
\]
We expand $\nabla \cdot ( \beta \ubold ) = \beta (\nabla \cdot \ubold) + \ubold
\cdot \nabla \beta$ and note that for the equality to hold we would
need
\[
\frac{1}{\beta(z)} \ubold \cdot  \nabla \beta 
 =  \frac{1}{\Gamma_1 p_0} \ubold \cdot \nabla p_0 \enskip ,
\]
or 
\[
\frac{1}{\beta} w \beta^\prime =  \frac{1}{\Gamma_1 p_0} w p_0^\prime \enskip .
\]
Since we want this to hold for all $w$, we are left with 
\[
\frac{\beta^\prime}{\beta}  =  \frac{p_0^\prime}{\Gamma_1 p_0} \enskip .
\]
We integrate this up from $z=0:$
\[
\int_0^z \frac{\beta^\prime}{\beta}\,dz^\prime  =  
\int_0^z \frac{d (\ln(\beta))}{dz}\,dz^\prime  =  
\int_0^z (\frac{p_0^\prime}{\Gamma_1 p_0})\,dz^\prime
\]
so
\[
\ln(\beta(z)) - \ln(\beta(0)) =  \int_0^z (\frac{p_0^\prime}{\Gamma_1 p_0})\,dz^\prime \enskip ,
\]
or 
\[
\beta(z) = \beta(0) \exp \left ({\int_0^z (\frac{p_0^\prime}{\Gamma_1 p_0})\,dz^\prime} \right ) \enskip .
\]
We note that this also can be written as the recursive relationship
\[
\beta(z_k) = \beta(z_{k-1}) \exp \left ({\int_{z_{k-1}}^{z_k} (-\frac{\rho_0 g}{\Gamma_1 p_0})\,dz^\prime} \right )\enskip,
\]
exploiting the hydrostatic equilibrium of the base state.  This
equation is the one we use to numerically compute $\beta(z)$; we let
$\beta(0) = \rho_0(0)$.

\pagebreak

\clearpage

\clearpage

\clearpage

\clearpage

\clearpage

\clearpage

\clearpage

\clearpage

\clearpage

\end{document}